\documentclass[12pt,letterpaper]{article}

\usepackage[top=1in, bottom=1in, left=1in, right=1in]{geometry}
\usepackage{lscape,pdflscape}
\usepackage{alltt,amsmath,caption,color,epsfig,enumerate,enumitem,float,
  graphicx,hyperref,indentfirst,pdfpages,relsize,sectsty,setspace,subcaption,times}
\hypersetup{colorlinks, allcolors={black}}  
\hypersetup{pdfpagemode=UseNone} 
\usepackage[authoryear]{natbib}
\bibpunct{(}{)}{;}{a}{}{,}

\allsectionsfont{\normalfont\sffamily\bfseries}
\newcommand{\LOD}{\text{LOD}}
\renewcommand{\figurename}{\small Figure}

\begin{document}

\setstretch{2.0}

\vspace*{8mm}
\begin{center}

\textbf{\Large The dissection of expression quantitative
  trait locus hotspots}

\bigskip \bigskip \bigskip \bigskip

{\large Jianan Tian$^{*}$,
Mark P. Keller$^{\dagger}$,
Aimee Teo Broman$^{\ddagger}$,
Christina Kendziorski$^{\ddagger}$, \\[-4pt]
Brian S. Yandell$^{*, \S}$,
Alan D. Attie$^{\dagger}$,
Karl W. Broman$^{\ddagger,1}$}

\bigskip \bigskip

Departments of $^{*}$Statistics,
$^{\dagger}$Biochemistry,
$^{\ddagger}$Biostatistics and Medical Informatics,
and $^{\S}$Horticulture,
University of Wisconsin--Madison, Madison, Wisconsin
53706
\end{center}

\newpage

\noindent \textbf{Running head:} Dissecting eQTL hotspots

\bigskip \bigskip \bigskip

\noindent \textbf{Key words:} eQTL, pleiotropy, multivariate analysis,
data visualization, gene expression

\bigskip \bigskip \bigskip

\noindent \textbf{$^1$Corresponding author:}

\begin{tabular}{lll}
 \\
 \hspace{1cm} & \multicolumn{2}{l}{Karl W Broman} \\
 & \multicolumn{2}{l}{Department of Biostatistics and Medical Informatics} \\
 & \multicolumn{2}{l}{University of Wisconsin--Madison} \\
 & \multicolumn{2}{l}{2126 Genetics-Biotechnology Center} \\
 & \multicolumn{2}{l}{425 Henry Mall} \\
 & \multicolumn{2}{l}{Madison, WI 53706} \\
 \\
 & Phone: & 608--262--4633 \\
 & Email: & \verb|kbroman@biostat.wisc.edu|
\end{tabular}

\newpage

\section*{Abstract}

Studies of the genetic loci that contribute to variation in gene
expression frequently identify loci with broad effect on gene
expression: expression quantitative trait locus (eQTL) hotspots. We
describe a set of exploratory graphical methods as well as a formal
likelihood-based test for assessing whether a given hotspot is due to
one or multiple polymorphisms. We first look at the pattern of effects
of the locus on the expression traits that map to the locus: the
direction of the effects, as well as the degree of dominance. A second
technique is to focus on the individuals that exhibit no recombination
event in the region, apply dimensionality reduction (such as with linear discriminant
analysis) and compare the phenotype distribution in the non-recombinants to
that in the recombinant individuals: If the recombinant
individuals display a different expression pattern than the
non-recombinants, this indicates the presence of multiple causal polymorphisms.
In the formal likelihood-based test, we compare a two-locus model, with
each expression trait affected by one or the other locus, to a
single-locus model. We apply our methods to a large mouse intercross
with gene expression microarray data on six tissues.

\newpage

\section*{Introduction}

There is a long history of efforts to map the genetic loci (called
quantitative trait loci, QTL)
that contribute to variation in quantitative traits in experimental organisms,
particularly to learn about the etiology of disease
\citep{broman2001review, jansen2007quantitative}. But it remains
difficult to identify the genes underlying QTL
\citep{nadeau2000roads}. There has recently been much interest in
measuring gene expression in disease-relevant tissues in QTL
experiments, as a way to speed the process from QTL to gene
\citep{jansen2001genetical,albert2015}. The genetic control of gene expression
is itself of great interest.

Expression quantitative trait loci (eQTL) analysis tries to find the
genomic locations that influence variation in gene expression levels
(mRNA abundances). eQTL near the genomic location of the influenced
gene are called local eQTL, and eQTL far away from the
influenced gene are called \emph{trans}-eQTL. When a
genomic region influences the expression of many genes, the region is called a
\emph{trans}-eQTL hotspot.

eQTL hotspots have been observed in many genetic studies \citep[e.g.,][]{brem2002,yvert2003,schadt2003,chesler2005},
and they are of particular interest
because gene expressions mapping to the same location may indicate the
existence of a genetic regulator.

Batch effects (artifacts arising from technical or environmental
factors) are common in microarray experiments. This has led to the
development of a number of methods to control for underlying
confounding factors \citep{leek2007,kang2008,stegle2010,listgarten2010,gagnon2012,fusi2012}.
However, these methods generally cannot distinguish \emph{trans}-eQTL
hotspots from batch effects. There is some controversy about whether
\emph{trans}-eQTL hotspots are themselves artifacts and whether one should
control for them, as one does for batch effects, in eQTL analysis
\citep[e.g.,][]{breitling2008genetical,kang2008}. But in many cases,
the associations between genotype and expression phenotypes are
extremely strong (with LOD $>$ 100), which largely precludes the
possibility of a batch effect artifact, as the strength of association
between batch and genotype in the region would have to be even stronger.

In \citet{tian2015}, we considered a large mouse intercross
between the strains C57BL/6J (abbreviated B6) and BTBR $T^+$ \emph{tf}/J
(abbreviated BTBR), with gene expression microarray data on six
tissues (adipose, gastrocnemius muscle, hypothalamus, pancreatic
islets, kidney, and liver), and mapped a \emph{trans}-eQTL hotspot to a
298 kb region containing just three genes. This sort of fine-mapping
approach is meaningful only if the hotspot is due to polymorphisms in
a single gene.

This raises an important question about \emph{trans}-eQTL hotspots:
are they the result of polymorphisms in a single gene, or are there
multiple underlying genes? In other words, is there complete
pleiotropy, or are there multiple linked eQTL?  Methods for testing
pleiotropy versus tight linkage of multiple QTL
\citep{jiang1995multiple,knott2000multitrait} do not scale well to the
case of the very large number of expression traits that map to a
\emph{trans}-eQTL hotspot. We developed a likelihood-based test that
is a variation on the method of \citet{knott2000multitrait}, as well
as a number of exploratory data visualizations, to test whether
multiple eQTL underlie a hotspot.  We apply our approaches to the data
considered in \citet{tian2015}.

\clearpage
\section*{Methods}

We focus on the case of an intercross between two inbred strains, B
and R (these labels were chosen to match the strains used in the
application, below). We assume dense marker genotype data and
genome-wide gene expression phenotype data (such as from microarrays
or RNA-sequencing). We first perform a genome scan to identify
quantitative trait loci (QTL), considering each expression trait
individually. We use Haley-Knott regression \citep{haley1992} for this
purpose, for the sake of speed. For each expression trait and each
chromosome, we consider the location of the single-largest LOD score,
provided that it exceeds a significance threshold that adjusts for the
genome scan but not the search across expression traits.

We count the number of expression traits that show a \emph{trans}-eQTL within a sliding
window (e.g., of 10~cM) across the genome, and use peaks in these
counts to define \emph{trans}-eQTL hotspots. We then focus on one such
hotspot, and on the set of expression traits that map to an interval
centered at the peak count.

We then ask: Are the multiple expression traits that map to this
\emph{trans}-eQTL hotspot all affected by a common eQTL, or could
there be multiple causal polymorphisms in the region?  We have
developed a set of exploratory data visualizations to address this
question, as well as a formal likelihood-based test. We prefer to
exclude expression traits whose genomic position is near (or even on
the same chromosome as) the hotspot
of interest, as these may be driven by separate, local-eQTL.

\subsection*{Exploratory data visualizations}

We first consider the pattern of effects of the locus on the
expression traits that map to the region: the direction of the
effects, as well as the degree of dominance. We use a pair of data
visualizations: first, a plot of the signed LOD score
(with the sign taken from the estimated additive effect) versus the
estimated eQTL location, for each expression trait analyzed
separately. That is, for each expression trait, we find the largest
LOD score on the chromosome, multiply it by $\pm$1, according to the
sign of the estimated additive effect of the locus, and plot this
signed LOD score versus the location at which that maximum LOD score
was attained. If there are two nearby loci with effects in opposite
directions, they may be revealed by this plot.

In addition, we plot the estimated dominance effect against the
estimated additive effect, for each expression trait. Let B and R
denote the two alleles in the cross, and let $\hat{\mu}_{\text{BB}}$,
$\hat{\mu}_{\text{BR}}$ and $\hat{\mu}_{\text{RR}}$ denote the average
expression levels for genotypes BB, BR, and RR, respectively.  We
estimate the additive effect as half the difference between the two
homozygotes, $\hat{a} =
(\hat{\mu}_{\text{RR}}-\hat{\mu}_{\text{BB}})/2$, and the dominance
effect as the difference between the heterozygote and the midpoint
between the two homozygotes, $\hat{d} =
\hat{\mu}_{\text{BR}}-(\hat{\mu}_{\text{BB}}+\hat{\mu}_{\text{RR}})/2$.
We then plot $\hat{d}$ vs $\hat{a}$ for all expression traits mapping
to the hotspot. If there are two nearby loci with different
inheritance patterns (e.g., one has additive allele effects and the
other has an allele that is dominant), they may be revealed by this
plot.

As a second technique, we consider the individuals that have no
recombination event in the region. For these individuals, we know
their eQTL genotype. We apply linear discriminant analysis
\citep[LDA,][Ch. 4]{hastie2009} to the top 100 traits with the largest LOD scores
and make a scatterplot of the first and
second linear discriminants; this should show three distinct clusters
(or, for a fully dominant locus, two clusters). We calculate the
linear discriminants for individuals that show a recombination event
in the region, and add them as points to the plot. If the recombinant
individuals fall within the clusters defined by the non-recombinants,
this is consistent with there being a single causal locus. If, on the
other hand, the recombinants look distinctly different from the
non-recombinants, then multiple polymorphisms are indicated.

The basic idea underlying this visualization is that the
non-recombinants can be used to derive an estimate
of the conditional distribution of the multivariate expression
phenotype, given the eQTL genotype. We use LDA as a
dimension-reduction technique. The goal of the visualization is to
compare the expression pattern in the recombinants and
non-recombinants. If there is a single eQTL, the recombinants should
look no different from the non-recombinants; if there is a difference,
we can conclude that there are multiple eQTL.

\subsection*{Formal statistical test}

To formally assess evidence of multiple linked loci versus complete
pleiotropy at a \emph{trans}-eQTL hotspot, we developed a
likelihood-based test to compare the null hypothesis of a single eQTL
affecting all expression traits to the hypothesis of two eQTL, with
each expression trait affected by one or the other eQTL (but not
both). The approach can handle only a limited number of expression
traits, and so we focus on the 50 traits with largest LOD scores (when
considered individually) in the interval centered at the
hotspot.

We assume that the traits follow a multivariate normal distribution,
conditional on eQTL genotype, and apply the multivariate QTL analysis
method of \citet{knott2000multitrait}. For a given QTL model, we have $Y = X
\beta + E$, where $Y$ is an $n \times p$ matrix of phenotypes,
with $n$ as the number of F$_2$ individuals and $p$ as the number of traits,
$X$ is an $n \times q$ matrix of covariates (including additive covariates,
interactive covariates, genotype probabilities for the position under
investigation, and the interactive covariates times the genotype
probabilities), and $\beta$ is a $q \times p$ matrix of
coefficients. We obtain $\hat{\beta} = (X'X)^{-1} X' Y$, calculate the
matrix of estimated residuals $\hat{E} = Y - X\hat{\beta}$, and calculate the residual
sum of squares matrix $\text{RSS} = \hat{E}'\hat{E}$. The LOD score is
$\frac{n}{2} \log_{10} \{|\text{RSS}_0| / |\text{RSS}|\}$, where
$|\text{RSS}|$ denotes the determinant of the RSS matrix, and
$\text{RSS}_0$ is the residual sum of squares matrix for the null
model (with additive covariates, but no genotype
probabilities or interactive covariates).

We perform a QTL scan over the
interval; at each putative QTL location, denoted $\lambda$, we
calculate the LOD score, $\LOD_1(\lambda)$,
comparing this single-QTL model to the null model of
no QTL. Let $M_1 = \max_\lambda \LOD_1(\lambda)$.

We compare this to a two-QTL model, in which each expression trait is affected by
one or the other QTL but not both. In principle, one would need to
consider, with $p$ expression traits, $2^{p-1}$ possible assignments of
the expression traits to the left and right QTL. This is a
prohibitively large number, and so we make an approximation: we sort
the expression traits according to their estimated QTL location, when
considered individually, and we consider only the $p-1$ cut-points of this
list. We randomly order any expression traits that map to the same
position. For each cut-point, we perform a two-dimensional scan over
possible two-QTL models, and calculate
$\LOD_2^{(c)}(\lambda_1, \lambda_2)$,
comparing the two-QTL model, with QTL at positions $\lambda_1$
and $\lambda_2$ and with the first $c$ expression traits affected by
the QTL at $\lambda_1$ and the last $p-c$ traits affected by the QTL
at $\lambda_2$. Let
$M_2^{(c)} = \max_{\lambda_1,\lambda_2} \LOD_2^{(c)}(\lambda_1, \lambda_2)$,
and let $M_2 = \max_c M_2^{(c)}$. The estimated cut-point is
$\hat{c} = \arg \max_c M_2^{(c)}$, and the estimated QTL positions are
$(\hat{\lambda}_1, \hat{\lambda}_2) = \arg \max_{\lambda_1, \lambda_2} \LOD_2^{(\hat{c})}(\lambda_1, \lambda_2)$.

As evidence for the presence of two QTL, we consider the
log$_{10}$ likelihood ratio, $\LOD_{2v1} = M_2 - M_1$.

The exhaustive two-dimensional scan of
$\LOD_2^{(c)}(\lambda_1,\lambda_2)$ is computationally intensive. We
can accelerate this calculation by iteratively searching for the
maximum on each of the two dimensions. There is no guarantee that this
will converge to the overall maximum, especially when there are
multiple modes in the two-QTL likelihood surface, but in practice
we've found this algorithm works well.  As a starting point of this
iterative search, we can use either the estimated QTL location under
the single-QTL model, or a randomly selected position.

\textbf{Statistical significance}:
To assess the statistical significance of the result, we need an
approximation of the distribution of the test statistic under the null
hypothesis of a single QTL. We consider two approaches: a parametric
bootstrap, and a stratified permutation test.

In the parametric bootstrap, we simulate new phenotype data using the
estimated single-QTL model. In the stratified permutation test, we
randomly permute the rows in the phenotype data relative to the rows in the genotype data,
within each QTL genotype group. When there are unmeasured genotypes at
the inferred QTL, we infer the QTL genotype, for each individual,
to be that with maximum probability, conditional on the observed
marker data. These conditional QTL genotype probabilities are
calculated by a hidden Markov model \citep[App.\ D]{broman_sen}.

For each procedure, we generate 1000 data sets, perform the full
likelihood analysis (the scan for the single-QTL model, and the
two-dimensional scan for the two-QTL model, for each possible
partition of the traits) and calculate the test statistic. The P-value
for the test is taken to be the proportion of simulated or
permuted data sets with a test statistic that is greater or equal to
the observed test statistic.

\textbf{Visualizations}:
To visualize the results for the two-QTL model, we plot profile LOD
curves for the left and right QTL, using the estimated
cut-point, $\hat{c}$, for the expression traits into those mapping to the left QTL
and those mapping to the right QTL. For the left QTL, we plot the
slice $\LOD_2^{(\hat{c})}(\lambda_1, \hat{\lambda}_2)$ against
$\lambda_1$, for varying
values of $\lambda_1$. Similarly, for the right QTL, we plot
$\LOD_2^{(\hat{c})}(\hat{\lambda}_1, \lambda_2)$ against $\lambda_2$,
for varying values of $\lambda_2$.
These sorts of profile LOD curves follow from an innovation of \citet{zeng2000}.

As a further diagnostic plot, we consider the statistic
$\LOD_{2v1}^{(c)} = M_2^{(c)} - M_1$, as a function of the cut-point, $c$.
This is evidence for two vs one QTL,
for a given cut-point, of the expression traits, into those that map
to the left QTL and those that map to the right QTL.
This displays the evidence for two vs one QTL as well as the
evidence for a particular split of the expression traits.

\clearpage

\section*{Application}

To illustrate our methods for the dissection of \emph{trans}-eQTL
hotspots, we consider a large mouse F$_2$ intercross \citep{tian2015}
with gene expression microarray data on six tissues.
The data are available at the QTL Archive, now part of the Mouse
Phenome Database: \\
\href{http://phenome.jax.org/db/q?rtn=projects/projdet&reqprojid=532}{
\tt \small http://phenome.jax.org/db/q?rtn=projects/projdet\&reqprojid=532}

\subsection*{Materials}

The experiment was carried out in order to identify genes and pathways
that contribute to obesity-induced type II diabetes. Greater than
500 offspring were generated from an F$_2$ intercross between
diabetes-resistant (C57BL/6J, abbreviated B6) and diabetes-susceptible
(BTBR \emph{T}$^{+}$\emph{ tf}/J, abbreviated BTBR) mouse strains.
All mice were genetically obese through introgression of the leptin
mutation (\emph{Lep$^{ob/ob}$}) and were sacrificed at 10 weeks of
age, the age when essentially all BTBR ob/ob mice are diabetic.

Mice were genotyped with the 5K GeneChip (Affymetrix). After data
cleaning, there were 519 F$_2$ mice genotyped at 2,057 informative
markers.  Gene expression was assayed with custom two-color ink-jet
microarrays manufactured by Agilent Technologies (Palo Alto, CA).  Six
tissues from each F$_2$ mouse were used for expression profiling:
adipose, gastrocnemius muscle (abbreviated gastroc), hypothalamus
(abbreviated hypo), pancreatic islets (abbreviated islet), kidney, and
liver. Tissue-specific mRNA pools were used for the
reference channel, and gene expression was quantified as the ratio of
the mean log$_{10}$ intensity (mlratio). For further details, see
\citet{keller2008}. In the final data set, there were 519 mice with
gene expression data on at least one tissue (487 for adipose, 490 for
gastroc, 369 for hypo, 491 for islet, 474 for kidney, and 483 for
liver). The microarray included 40,572 total probes; we focused on the
37,797 probes with known location on one of the autosomes or the X
chromosome.

\subsection*{QTL analysis}

For QTL analysis, we first transformed the gene expression measures
for each microarray probe in each of the six tissues to normal
quantiles, taking $\Phi^{-1}[(R_i - 0.5)/n]$, where $\Phi$ is the
cumulative distribution function for the standard normal distribution
and $R_i$ is the rank in $\{1, \dots, n\}$ for mouse $i$. We then
performed single-QTL genome scans, separately for each probe in each
tissue, by Haley-Knott regression \citep{haley1992} with microarray
batch as an additive covariate and with sex as an interactive
covariate (i.e. allowing the effects of QTL to be different in the two
sexes).  Calculations were performed at the genetic markers and at a
set of pseudomarkers inserted into marker intervals, selected so that
adjacent positions were separated by $\le$~0.5~cM. We calculated
conditional genotype probabilities, given observed multipoint marker
genotype data, using a hidden Markov model assuming a genotyping error
rate of 0.2\%, and with genetic distances converted to recombination
fractions with the Carter-Falconer map function \citep{carter1951}.
Calculations were performed with R/qtl \citep{broman2003Rqtl}, an
add-on package to the general statistical software R \citep{RCore}.

For each probe in each tissue, we focused on the single largest LOD
score peak on each chromosome, and on LOD score peaks $\ge$ 5
(corresponding to genome-wide significance at the 5\% level, for a
single probe in a single tissue, determined by computer simulations
under the null hypothesis of no QTL).

\begin{figure}[ht]
  \centering
  \includegraphics[width=\textwidth]{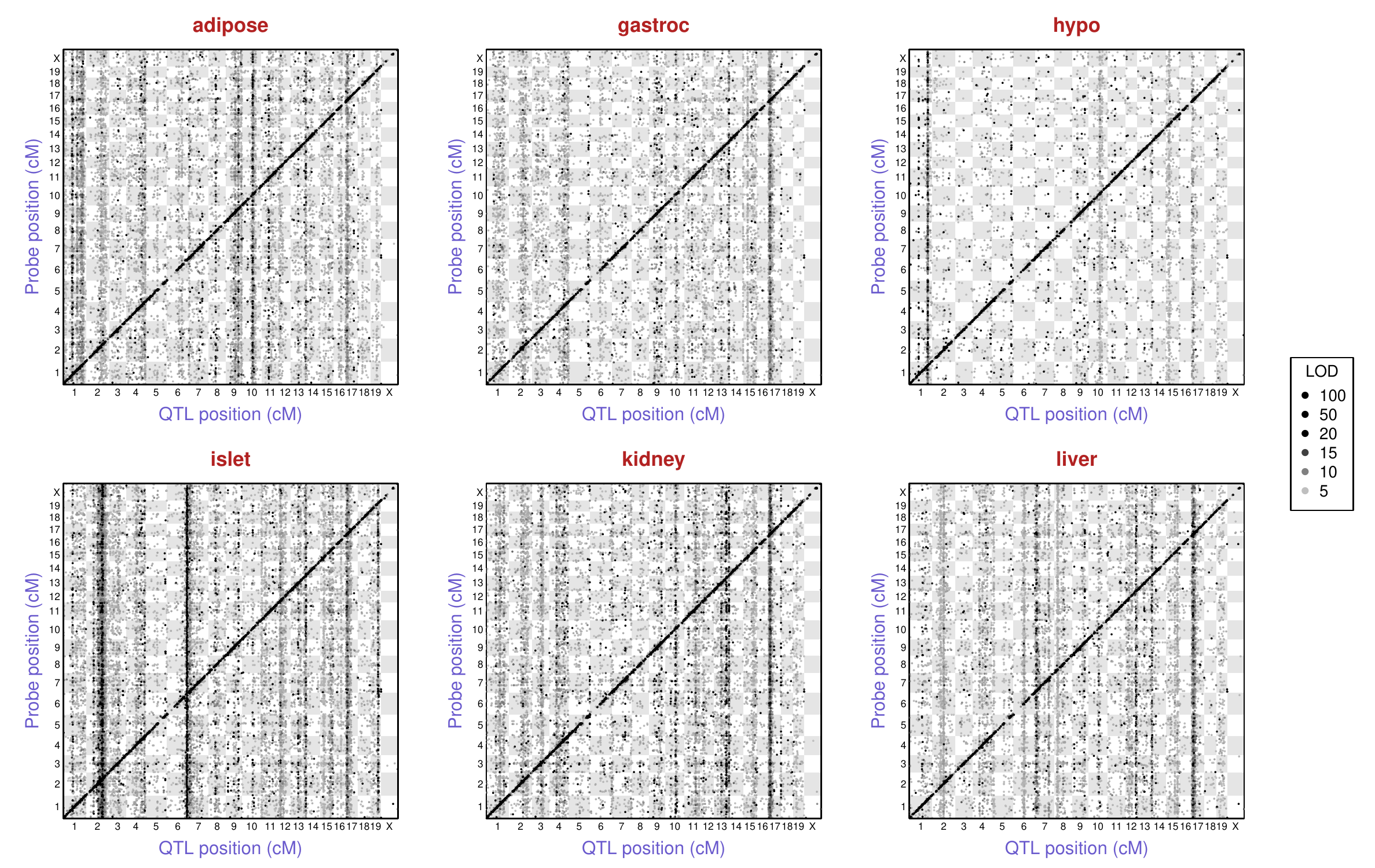}
  \caption{\small Inferred eQTL with LOD $\ge$ 5, by tissue. Points
  correspond to peak LOD scores from single-QTL genome scans with each
  microarray probe with known genomic position. The y-axis is the
  position of the probe and the x-axis is the inferred QTL position.
  Points are shaded according to the corresponding LOD score, though
  we threshold at 100: all points with LOD $\ge$ 100 are black.
  \citep[A version of this figure appeared as Figure 1 in][]{tian2015} \label{fig:eqtl}}
\end{figure}

The inferred eQTL for all genes with LOD $\ge$ 5 are displayed in
Figure~\ref{fig:eqtl}, with the y-axis corresponding to the genomic
position of the microarray probe and the x-axis corresponding to the
estimated eQTL position. As expected, we see a large number of
local-eQTL along the diagonal for each tissue-specific panel. These
local-eQTL correspond to genes for which expression or mRNA abundance
is strongly associated with genotype near their genomic position.

In addition to the local-eQTL, there are a number of prominent
vertical bands: genomic loci that influence the expression of genes
located throughout the genome. These are the \emph{trans}-eQTL
hotspots. Overall, we detected many more \emph{trans}-eQTL than
local-eQTL. The \emph{trans}-eQTL hotspots can show either remarkable
tissue specificity or be observed in multiple tissues. For example, a
locus near the centromere of chromosome~17, at 11.7~cM, shows effects
in all tissues. In contrast, the \emph{trans}-eQTL hotspot located at
the distal end of chromosome 6 was only observed in pancreatic
islets.

To define \emph{trans}-eQTL hotspots of potential interest, we focused
on a more conservative threshold for eQTL, of LOD $\ge$ 10. We further
excluded local-eQTL, defined here to be those for which the distance
between the gene's genomic position and its inferred eQTL position was
$<$ 10~cM. We then counted the number of expression traits with a
\emph{trans}-eQTL in a sliding interval of length 10~cM (Figure~S1).

For each \emph{trans}-eQTL of interest, we widened the interval to be
considered beyond that initial 10~cM window, to consider the interval
in which the count of expression traits with eQTL was $>$ 50, and then
padded this further by adding 5~cM on either end.

We will focus on a set of six hotspots:
adipose chromosome~1 at 39~cM, adipose chromosome~10 at 48~cM,
islet chromosome~2 at 75~cM,
islet chromosome~6 at 91~cM,
kidney~chromosome 13 at 68~cM,
and liver chromosome~17 at 18~cM.
Results for additional hotspots are
displayed in File~S1.

\subsection*{Visualization of QTL effects}

\begin{figure}
  \centering
  \includegraphics[height=7.8in]{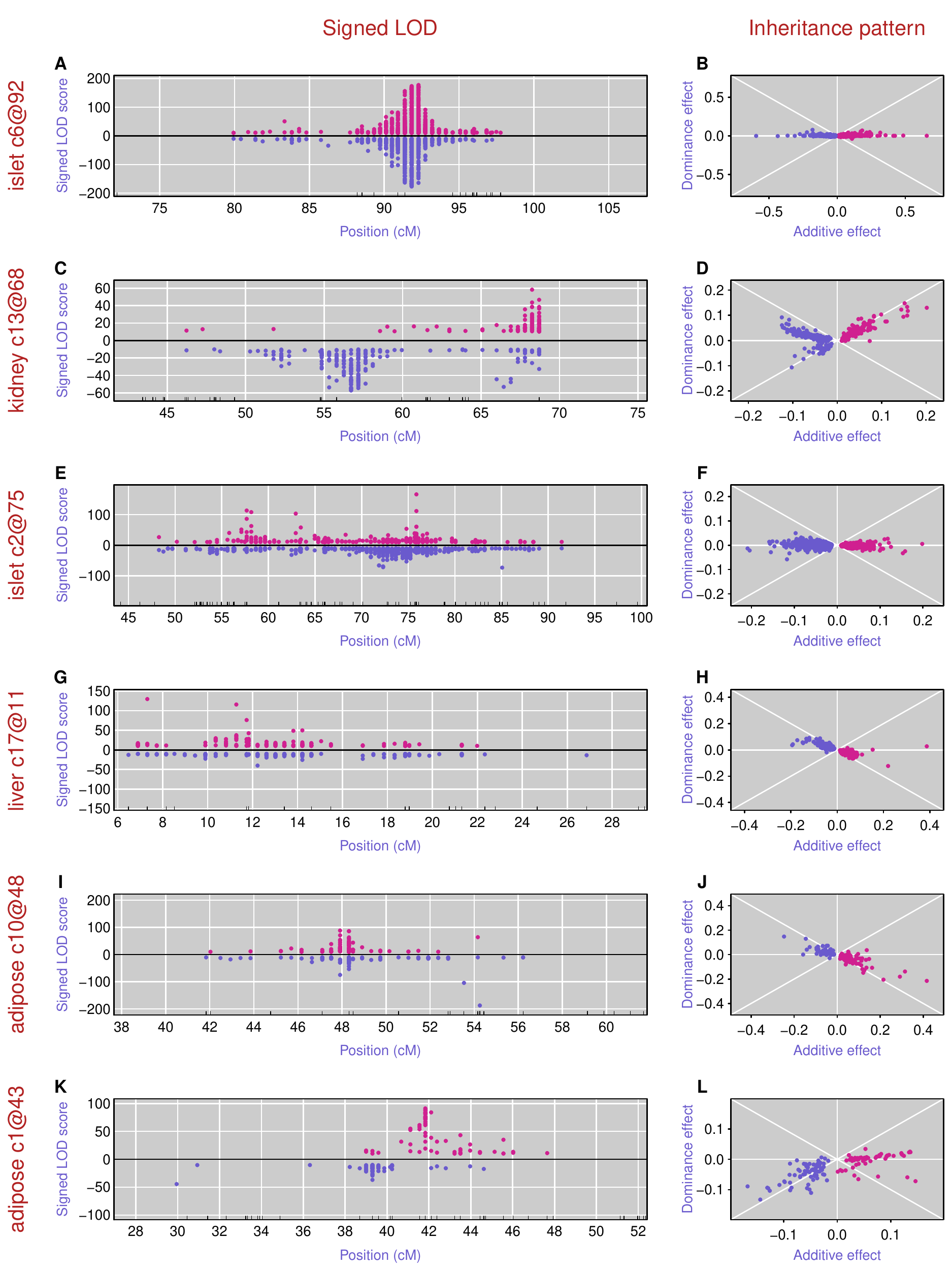}
  \caption{\small Visualizations of the QTL effects on the
    multiple expression traits that map with LOD $\ge$ 10 to a \emph{trans}-eQTL
    hotspot. Each row is a hotspot. The left panels are scatterplots
    of signed LOD scores (with positive values indicating that the
    BTBR allele is associated with larger average gene expression and negative
    values indicating that the B6 allele is associated with larger
    average gene expression) versus the estimated QTL location. Each
    point is a single expression trait. Tick marks at the bottom
    indicate the locations of the genetic markers. The right panels
    are scatterplots of the estimated dominance effects versus the
    estimated additive effects. \label{fig:effects}}
\end{figure}

We first consider the estimated effects of a locus on the expression
traits that map to the region (Figure~\ref{fig:effects}).  In the left
panels, we display the signed LOD score (with positive values
indicating that the BTBR allele is associated with larger average
expression and negative values indicating that the B6 allele is
associated with larger average expression) versus the estimated eQTL
location. In the right panels, we plot the estimated dominance effect
versus the estimated additive effect for all transcripts mapping to
the hotspot. The key value, in these visualizations, is for the case
that two linked QTL show distinct inheritance patterns.

The islet chromosome~6 hotspot, at 92~cM, shows approximately equal
numbers of expression traits for which the BTBR allele causes an
increase or decrease in gene expression (Figure~\ref{fig:effects}A),
and the allele effects are approximately additive
(Figure~\ref{fig:effects}B), with estimated dominance effect near
0. These results are consistent with there being a single QTL.
In \citet{tian2015}, this locus was resolved to a 298~kb interval
containing just three genes, with good evidence for \emph{Slco1a6\/}
as the causal gene.

The kidney chromosome~13 hotspot, at 68~cM, shows clear evidence for two
QTL. In Figure~\ref{fig:effects}C, we see that for expression traits
mapping to $\sim$57~cM, the BTBR allele is associated with a decrease
in expression, while for traits mapping to $\sim$68~cM, the BTBR
alleles is predominantly associated with an increase in expression,
though with some traits having effects in the opposite direction. From
Figure~\ref{fig:effects}D, we can infer that, for the traits mapping
to $\sim$57~cM, the B6 allele is nearly dominant ($d \approx -a$,
along the line with slope --1), while for the traits mapping to
$\sim$68~cM, the BTBR allele is dominant ($d \approx a$, along the
line with slope +1).

The islet chromosome~2 hotspot, at 75~cM, shows expression traits with
high LOD scores across a broad region (Figure~\ref{fig:effects}E).
For traits mapping to 70--75~cM, the B6 allele is associated with
increased expression, while for traits mapping to 55--60~cM, the
effect is in the opposite direction. The allele effects are nearly
additive for all expression traits (Figure~\ref{fig:effects}F).

The liver chromosome~17 hotspot, at 11~cM, has approximately equal
numbers of traits with effects in each direction
(Fig~\ref{fig:effects}G), and the B6 alleles appears to be nearly
dominant in most cases (Figure~\ref{fig:effects}H).
The adipose chromosome~10 hotspot, at 48~cM, is similar, with effects in
both directions (Figure~\ref{fig:effects}I) and with the B6 allele
being nearly dominant (Figure~\ref{fig:effects}J).

The adipose chromosome~1 hotspot, at 43~cM, again shows evidence for
two QTL. For the expression traits mapping to 38--40~cM, the B6 allele
is associated with increased expression (Figure~\ref{fig:effects}K),
but the BTBR allele appears dominant (Figure~\ref{fig:effects}L). For
traits mapping to 42-46~cM, however, the BTBR allele is associated
with increased expression and the allele effects appear additive.

In summary, for two out of these six hotspots, these visualizations of
the estimated QTL effects provide good evidence for two QTL. In one
case (kidney chromosome~13), the two QTL are well-separated,
but in the other case (adipose chromosome~1), the two loci are
tightly linked.

\subsection*{Comparison of recombinants and non-recombinants}

\begin{figure}
  \centering
  \includegraphics[height=8.2in]{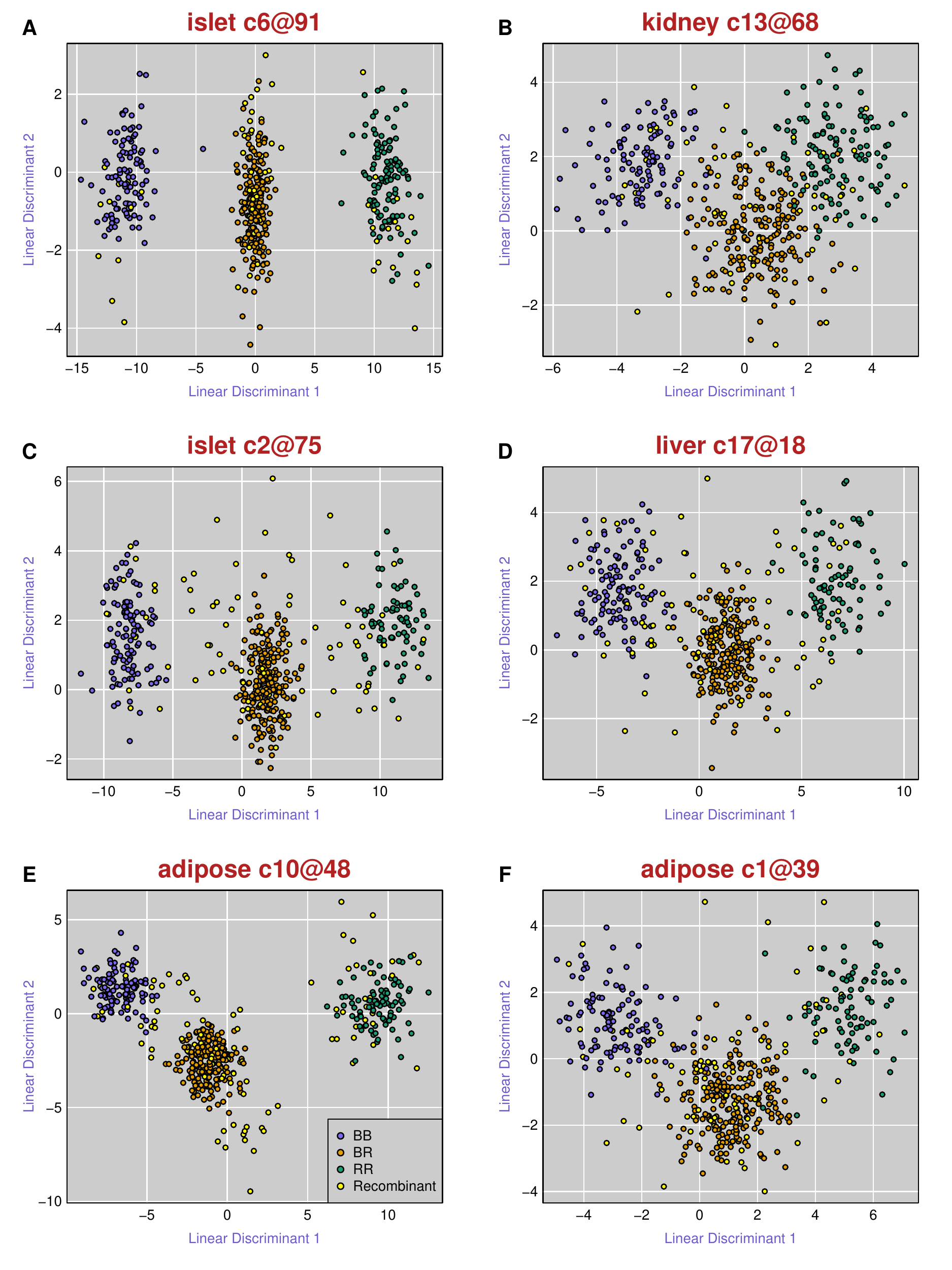}
  \caption{\small Scatterplots of the first two linear discriminants from
    application of linear discriminant analysis to the 100 expression
    traits that map to the region with the highest LOD score, with mice
    that show no recombination event in the region of a
    \emph{trans}-eQTL hotspot. Blue, orange, and green points
    correspond to the non-recombinant mice with genotype BB, BR, and
    RR, respectively, at the eQTL. Yellow points correspond to
    recombinant mice.\label{fig:lda}}
\end{figure}

Our second graphical technique is to consider the individuals
exhibiting no recombination event in the region of a \emph{trans}-eQTL
hotspot (for these individuals, we know their eQTL genotype), apply
linear discriminant analysis (LDA) using the top 100 expression traits
that map to the region, and make a scatterplot of the first two linear
discriminants. Superposing points for the recombinant individuals, we
can make a direct comparison of the recombinants and
non-recombinants (Figure~\ref{fig:lda}). If there is a single eQTL in the region, the
recombinant individuals should reside within the clusters defined by
the non-recombinant individuals. If the recombinant individuals appear
different from the non-recombinants, this indicates the presence of a
second QTL.

For the islet chromosome~6 hotspot, the non-recombinant mice form
three distinct clusters, and the recombinant mice (in yellow) fit
reasonably well within those clusters (Figure~\ref{fig:lda}A). This is
consistent with there being a single eQTL.

For the islet chromosome~2 (Figure~\ref{fig:lda}C) and
adipose chromosome~10 (Figure~\ref{fig:lda}E) hotspots, the
non-recombinant mice again form tight clusters, but the recombinant
mice fall clearly outside those clusters. This is evidence for the
presence of more than one eQTL.

In the other three cases, kidney chromosome~13 (Figure~\ref{fig:lda}B),
liver chromosome~17 (Figure~\ref{fig:lda}D), and adipose chromosome~1
(Figure~\ref{fig:lda}F), the clusters of non-recombinants are not so
tight, and the recombinants are not obviously different from the
recombinants. However, for the liver chromosome~17 hotspot
(Figure~\ref{fig:lda}D), one might make the case that the majority of recombinants are at
the boundaries between the clusters, and so multiple eQTL may be indicated.

\begin{figure}
  \centering
  \includegraphics[width=\textwidth]{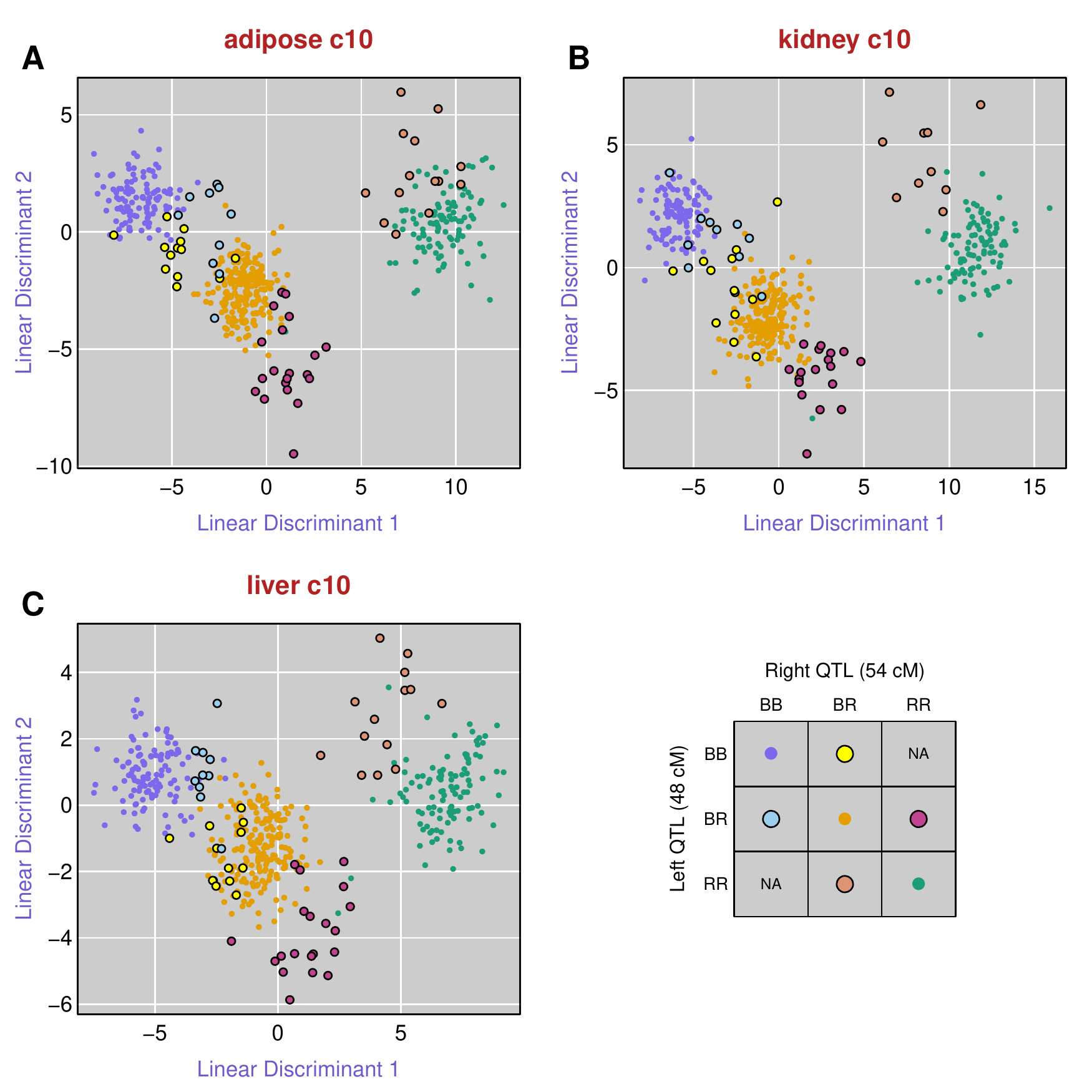}
  \caption{\small Scatterplots of the first two linear discriminants, as in
    Figure~\ref{fig:lda}E, for the \emph{trans}-eQTL hotspot on
    chromosome 10, here considering three tissues: adipose, kidney,
    and liver. Points correspond to mice, and they are colored
    according to their two-locus genotypes, for the inferred two QTL
    model, with one locus at $\sim$48~cM and the other at
    $\sim$54~cM.\label{fig:lda_2qtl}}
\end{figure}

Returning to the adipose chromosome~10 hotspot
(Figure~\ref{fig:lda}E), note how the
recombinant mice form tight clusters that are
distinct from the non-recombinants. If there are two eQTL in the
region, perhaps these clusters correspond to different two-locus
recombinant genotypes? Through the fit of a two-QTL
model (in the next section), we estimate the two QTL to be at 48 and
54~cM. If we color the points by the two-locus genotypes for these two
positions, we see that the clusters of non-recombinants do share a common
two-locus genotype (Figure~\ref{fig:lda_2qtl}A).

This chromosome~10 hotspot also shows effect in kidney and liver, and
so we applied this technique for this same region, with expression
data for these tissues (Figures~\ref{fig:lda_2qtl}B and
\ref{fig:lda_2qtl}C). The three tissues give consistent results.
Mice that are heterozygous at one QTL and homozygous BB at the other
(yellow and light blue)
sit between the non-recombinants that are homozygous BB (dark blue) and those that
are heterozygous (orange). Mice that are homozygous RR at the left QTL and
heterozygous at the right QTL (brown) sit above the non-recombinant RR
mice (green),
while mice that are heterozygous at the left QTL and homozygous RR at
the right QTL (red) sit below the non-recombinant heterozygotes (orange).

There is one green point (non-recombinant RR) sitting among the
red points (BR at the left QTL and RR at the right QTL). This mouse
(with ID 3117) has
a recombination event just to the left of the left QTL; if we moved that
QTL slightly to the left, it would become a red point (BR at the left
QTL and RR at the right QTL). In principle, a series of graphs of
this form, with varying locations for the left and right QTL, could be
used to define the QTL intervals in the context of this two-QTL model.

There is one additional green point among the red points in
Figure~\ref{fig:lda_2qtl}C (liver). This mouse (with ID 3317) sits
at the center of the cluster of green points in
each of Figures~\ref{fig:lda_2qtl}A and \ref{fig:lda_2qtl}B, and shows no
recombination event in the region of these two QTL.

This example illustrates that consideration of the two-locus
genotypes can help to strengthen evidence for two loci underlying a
\emph{trans}-eQTL hotspot. However, as we will describe below, in this
particular case, the right eQTL appears to affect just three of the expression
traits. Just one single trait, if affected by a separate locus, can
have a great deal of leverage on these sorts of plots.

\subsection*{Formal tests for two QTL}

\begin{figure}
  \centering
  \includegraphics[height=7.4in]{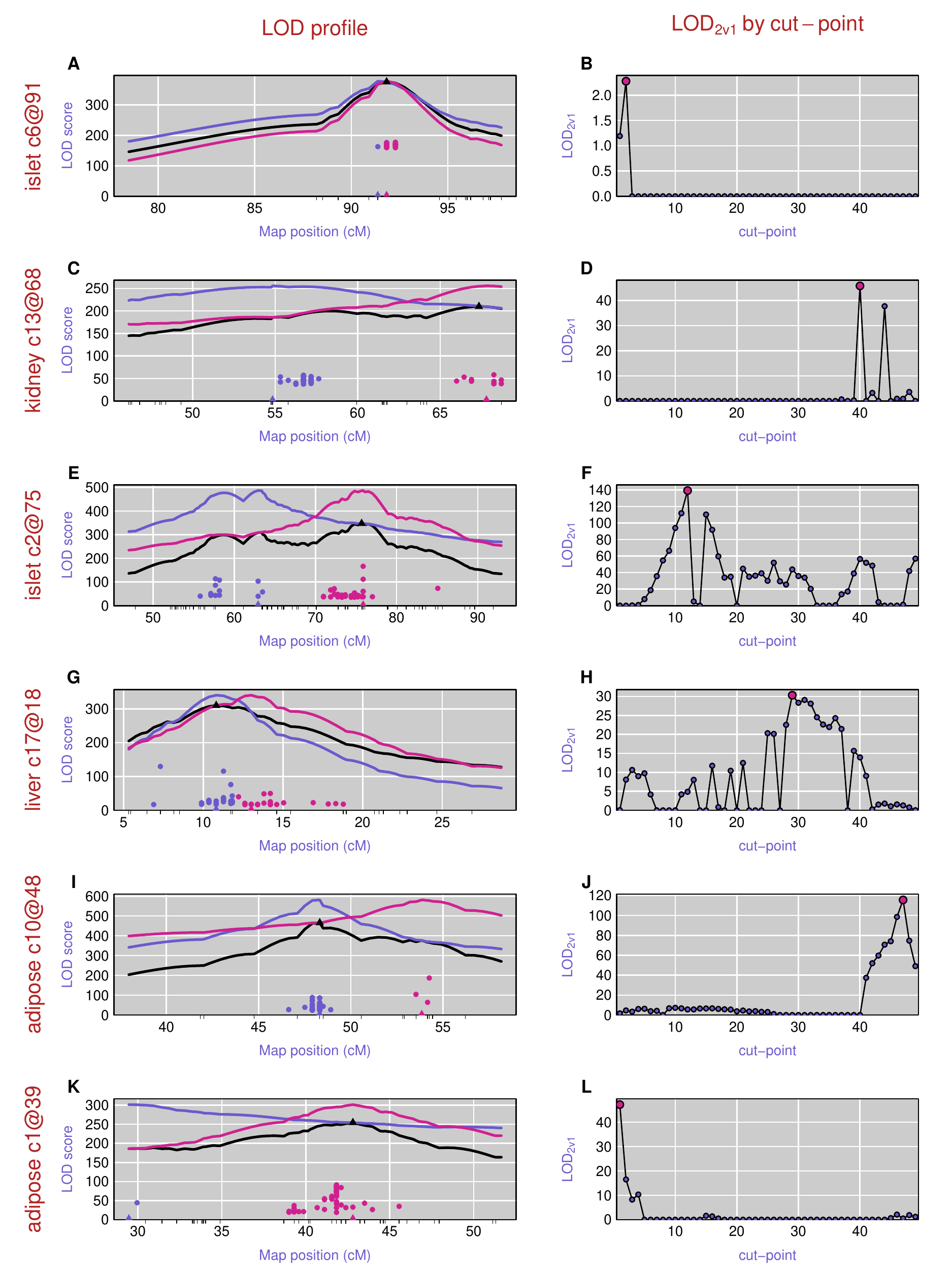}
  \caption{\small Results of a test of one versus two QTL at a
    \emph{trans}-eQTL hotspot, considering the top 50 traits, in terms
    of LOD score, that map to the region. Each row is a hotspot. In
    the left panels, the black curve is the LOD curve for the
    single-QTL model, with estimated QTL location indicated by a black
    triangle. The blue and pink curves are profile LOD curves for the
    left and right QTL, respectively, for the estimated two-QTL
    model(with the estimated cut-point). Points indicate the LOD score and estimated QTL position
    for the 50 expression traits, analyzed separately. The points are
    colored according to whether they are estimated to be affected by
    the left QTL (blue) or right QTL (pink). The right panels show the
    $\LOD_{2v1}^{(c)}$ score, indicating evidence for two versus one
    QTL, for each possible cut-point, $c$, of the list of expression
    traits into those that map to the left and those that map to the
    right QTL.\label{fig:formal}}
\end{figure}

To supplement these visualization techniques, we developed a formal
statistical test for whether a \emph{trans}-eQTL hotspot harbors one
vs two eQTL. The results of this approach, for the six hotspots under
consideration, are displayed in Figure~\ref{fig:formal}.

Let's begin by considering the kidney chromosome~13 hotspot
(Figures~\ref{fig:formal}C and \ref{fig:formal}D). For these
multivariate likelihood analyses, we focus on the top 50 expression
traits mapping to the region, in terms of their LOD scores when
considered individually. In the left panel
(Figure~\ref{fig:formal}C), the black curve is the LOD curve for the
multivariate QTL analysis with a single-QTL model. The estimated QTL
location is at 67.4~cM. The blue and pink curves are LOD profiles for
the estimated two-QTL model, for which the estimated QTL locations are
at 54.8 and 67.8~cM. The blue and pink points indicate the LOD scores
and estimated QTL locations for the individual expression traits, with
blue points affected by the left QTL and pink points affected by the
right QTL. The right panel (Figure~\ref{fig:formal}D) shows the
evidence for two versus one QTL as a function of the choice of
cut-point for the list of expression traits, into those affected by
the left and right QTL. The inferred cut-point has 40 traits affected
by the left QTL and 10 traits affected by the right QTL, and has a
LOD$_{2v1}$ score of 45.8, indicating very strong evidence for two
QTL, and for this particular cut-point.

The results for the islet chromosome~6 hotspot are displayed in
Figures~\ref{fig:formal}A and \ref{fig:formal}B. The inferred two-QTL
model has QTL at 91.4 and 91.8~cM, with only the two expression
traits affected by the left QTL. And LOD$_{2v1}$ = 2.3, indicating weak
evidence for two QTL.

The results for the islet chromosome~2 hotspot (Figures~\ref{fig:formal}E and
\ref{fig:formal}F) indicate strong evidence for two QTL, with
LOD$_{2v1}$ = 139. The estimated QTL are at 62.9 and 75.7~cM. The left
QTL is inferred to affect 12 of the 50 expression traits.

The liver chromosome~17 hotspot (Figures~\ref{fig:formal}G and
\ref{fig:formal}H) has strong evidence for two QTL, with LOD$_{2v1}$ =
30, and the estimated QTL locations at 10.8 and 13.0~cM. The choice of
cut-point of the expression traits is not so clear. We estimate 29 of
the expression traits are affected by the left eQTL, but a model with
32 traits affected by the left eQTL gives a similar likelihood.

For the adipose chromosome~10 (Figures~\ref{fig:formal}I and
\ref{fig:formal}J) and the adipose chromosome~1
(Figures~\ref{fig:formal}K and \ref{fig:formal}L) hotspots, the
evidence for two QTL is strong, but only three eQTL are inferred to be
affected by the right eQTL at the chromosome~10 hotspot, and only 1
trait is inferred to be affected by the left eQTL at the chromosome~1
hotspot. If we trim off these expression traits and apply the
procedure again, we find that, for the adipose chromosome~10 locus
(see Figure~S2), there is little evidence for more than one eQTL
affecting the remaining traits. Further, the analysis of two-locus
genotypes in the LDA plot in Figure~\ref{fig:lda_2qtl} is largely
driven by these three expression traits that map to 54~cM. Similarly,
if we trim off the first expression trait for the adipose chromosome~1
hotspot (Figure~S3), there is limited evidence for multiple QTL
affecting the remaining traits.

In summary, the formal statistical test provides strong evidence for
two eQTL in five of these six cases, but in two of the cases, the
majority of the traits are affected by a single eQTL.

\clearpage
\section*{Simulations}

To further assess the performance of the proposed likelihood-based
test for whether a \emph{trans}-eQTL hotspot harbors more than one
eQTL, we performed a set of computer simulation studies.

We generated 500 intercross offspring with 100
markers on a 100~cM chromosome, and then simulated $p$ = 10 or 40 traits,
with half of the traits affected by a QTL at 50~cM and
the other traits affected by a QTL 0 -- 20~cM away (at 50
-- 70~cM). We also considered an unbalanced case with 5 traits
affected by the left QTL and 35 traits affected by the right QTL.
We assumed additive allele effects, with the additive effect of each
QTL being $a$ = 0.1, 0.2,
0.3, 0.4, or 0.5.
Residual variation followed a normal distribution with mean 0 and
standard deviation 1, with traits conditionally independent given the
QTL genotypes.

We used 100 simulation replicates for each situation and calculated P-values
by a parametric bootstrap with 1000 simulation replicates.

\begin{figure}
  \centering
  \includegraphics[width=0.8\textwidth]{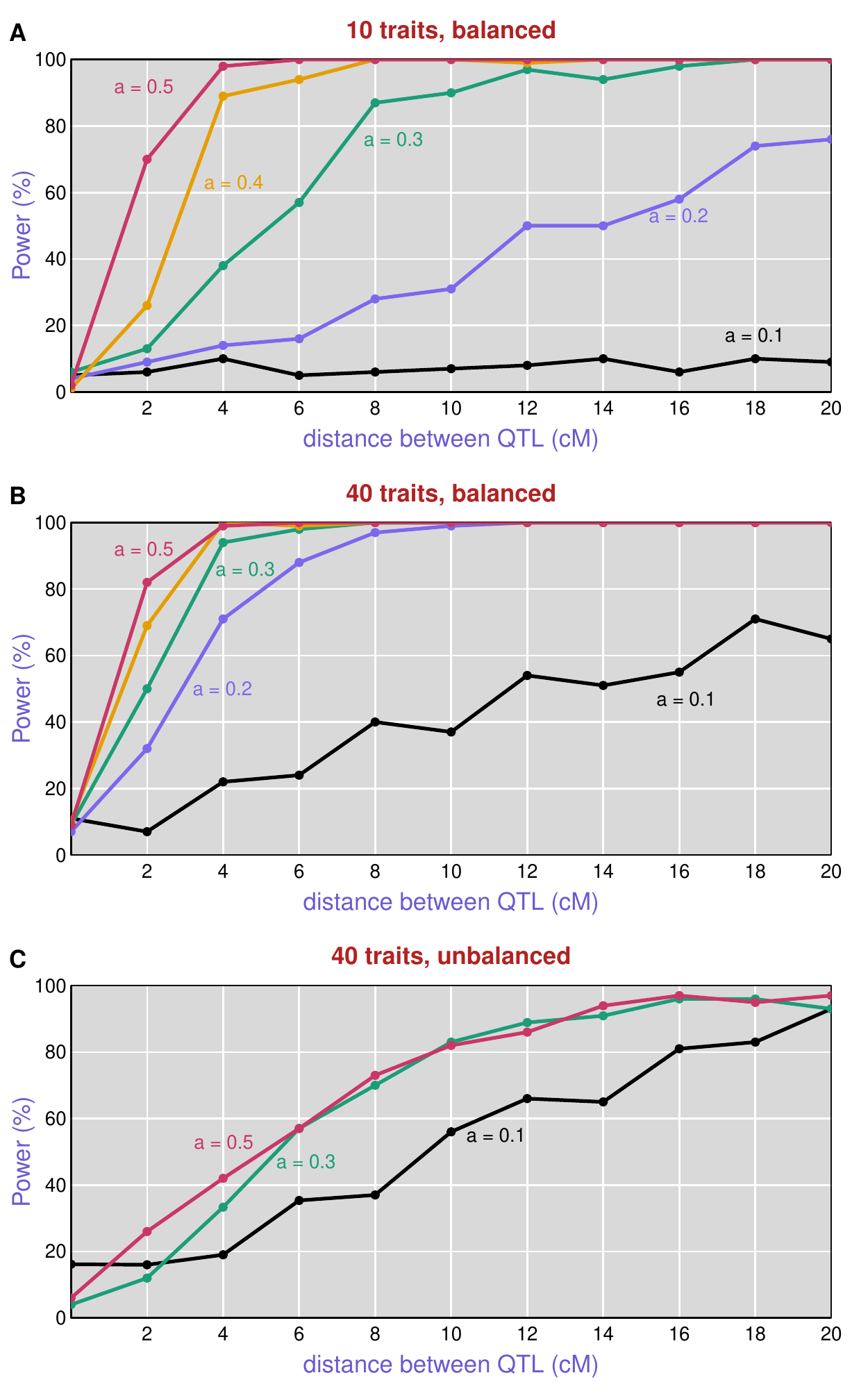}
  \caption{\small Power to detect two QTL as a function of the
    distance between the QTL, for varying QTL effects. (A) Ten traits,
    with each QTL affecting five traits. (B) 40 traits, with each QTL
    affecting 20, and (C) 40 traits, with the left QTL affecting five
    and the right QTL affecting 35.}

  \label{fig:power}
\end{figure}

The estimated power to detect two linked QTL, as a function of the
distance between the QTL, is shown in
Figure~\ref{fig:power}. When the QTL effect is smaller than 0.3, the power to distinguish two QTL within distance of 10~cM
is low for $p=10$. For any fixed effect, the power to
detect two QTL is higher for $p=40$ than for $p=10$.
When the QTL effect is larger than 0.4, the power to distinguish two
QTL separated by more than 5~cM
is almost 100\%.

The power to detect two QTL in the unbalanced case, with the left QTL affecting 5/40 traits,
(Figure~\ref{fig:power}C) is considerably lower than for the balanced
case (Figure~\ref{fig:power}B).

The case of distance = 0 corresponds to the null hypothesis of a
single QTL affecting all traits. The power in this case is the type I
error rate, and the parametric bootstrap method is seen to have
somewhat inflated type 1 error rate, relative to the nominal 5\%.

\clearpage
\section*{Discussion}

In this paper, we have proposed exploratory methods and a formal
inference method for dissecting \emph{trans}-eQTL hotspots. We applied
these approaches to data on a large mouse intercross with
gene expression microarray data on six tissues, and we performed
a simulation study to investigate the
performance of the formal inference method. Both
the exploratory methods and the formal inference method are helpful in
dissecting \emph{trans}-eQTL hotspots, and can give improved estimates
of the eQTL positions.

The exploratory
methods have the advantage of providing insight into the underlying
evidence for multiple eQTL: the multiple eQTL may show distinct inheritance patterns,
or the recombinant and non-recombinant individuals may show
differences in expression. However, while the visualization methods
can be strongly informative, they will not necessarily reveal the
presence of two eQTL, as the inheritance pattern of the two linked
eQTL may be the same, or the first two linear discriminants may not be
revealing of the difference between the recombinants and
non-recombinants.

In forming a multivariate test statistic, we chose to follow the
method of \citet{knott2000multitrait}, but other multivariate analysis of
variance (MANOVA) statistics could also be used, including Pillai's
trace, Lawley-Hotelling's test, and Roy's Lambda
\citep{anderson1958introduction}. Similarly, in the exploratory data
visualization based on linear discriminant analysis, other
dimension-reduction techniques could be used. A supervised (i.e.,
classification) method, which makes use of the known eQTL genotypes of
the non-recombinant individuals, is preferred.

The main issue in the formal statistical test is the choice of
expression traits, as we can't handle a very large number of
expression traits. Our choice to focus on the 50 traits with the
highest LOD score was arbitrary, and deserves further investigation.
Regularized methods \citep[see][Sec
  5.8]{hastie2009}, or a Bayesian approach, might
have an advantage in this context. Such approaches could also be used
to relax some of our modeling assumptions. For example, one might
consider a model with two eQTL, where each expression trait can be affected
by both.

We considered two methods to calculate p-values: a parametric
bootstrap, and a stratified permutation test.  The simulations to
investigate power used the parametric bootstrap, and showed somewhat
inflated type I error rate in the case. Moreover, neither approach
takes account of the selection of hotspots, which may introduce
further bias.

We considered a single tissue at a time. The joint consideration of
multiple tissues could provide additional power to dissect
\emph{trans}-eQTL hotspots that are in common across tissues.

We ignored the effects of eQTL elsewhere in the genome and considered
just one region in isolation. In doing so, the effects of any other
eQTL become part of the residual variation. Local-eQTL are a
particularly important case, as they are quite common and often have
large effect. Controlling for the effect of local-eQTL could give
better precision in the dissection of a \emph{trans}-eQTL hotspot.

We have implemented our methods in an R package \citep{RCore},
{\em qtlpvl}, available at
\href{https://github.com/jianan/qtlpvl}{\tt https://github.com/jianan/qtlpvl}.

\clearpage
\section*{Acknowledgments}

The authors thank two anonymous reviewers for comments to improve the
manuscript and Amit Kulkarni for providing annotation information
for the gene expression microarrays. This work was supported in part
by National Science Foundation grant DMS-12-65203 (to C.K.) and
National Institutes of Health grants GM074244 (to K.W.B.), GM070683
(to K.W.B.), DK066369 (to A.D.A), and GM102756 (to C.K.).

\clearpage
\bibliographystyle{genetics}
\renewcommand*{\refname}{\normalfont\sffamily\bfseries Literature Cited}
\bibliography{eqtl}

\newpage

\captionsetup[figure]{labelsep=quad}
\captionsetup[table]{labelsep=quad}

\renewcommand{\thefigure}{\textbf{S\arabic{figure}}}
\renewcommand{\figurename}{\textbf{Figure}}

\renewcommand{\thetable}{\textbf{S\arabic{table}}}
\renewcommand{\tablename}{\textbf{Table}}

\vspace*{8mm}
\begin{center}

\textbf{\Large The dissection of expression quantitative
  trait locus hotspots}

\bigskip \bigskip \bigskip \bigskip

\textbf{\Large SUPPLEMENT}

\bigskip \bigskip
\bigskip \bigskip

{\large Jianan Tian$^{*}$,
Mark P. Keller$^{\dagger}$,
Aimee Teo Broman$^{\ddagger}$,
Christina Kendziorski$^{\ddagger}$, \\[4pt]
Brian S. Yandell$^{*, \S}$,
Alan D. Attie$^{\dagger}$
Karl W. Broman$^{\ddagger,1}$}

\bigskip \bigskip

Departments of $^{*}$Statistics,
$^{\dagger}$Biochemistry,
$^{\ddagger}$Biostatistics and Medical Informatics,
and $^{\S}$Horticulture, \\[4pt]
University of Wisconsin--Madison, Madison, Wisconsin
53706
\end{center}

\clearpage

\addtocounter{figure}{-6}

\begin{figure}[p]
\centerline{\includegraphics[height=0.92\textheight]{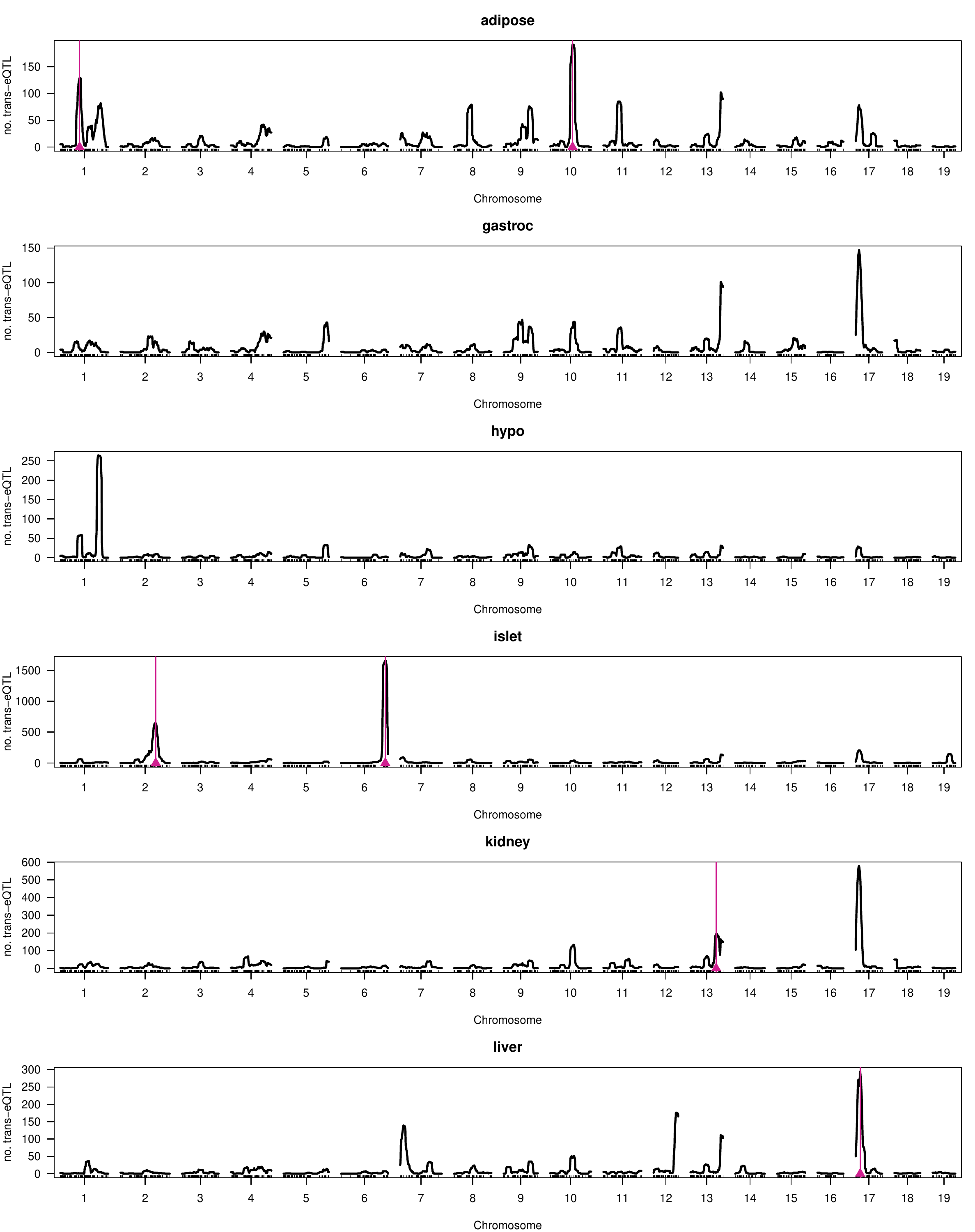}}

\caption{Number of expression traits with LOD $\ge$ 10 in a 10 cM
  sliding window across the genome. Red triangles indicate the six
  \emph{trans}-eQTL hotspots used as examples in Figures 2, 3,
  and 5.}
\end{figure}

\clearpage

\begin{figure}[p]
\centerline{\includegraphics[width=\textwidth]{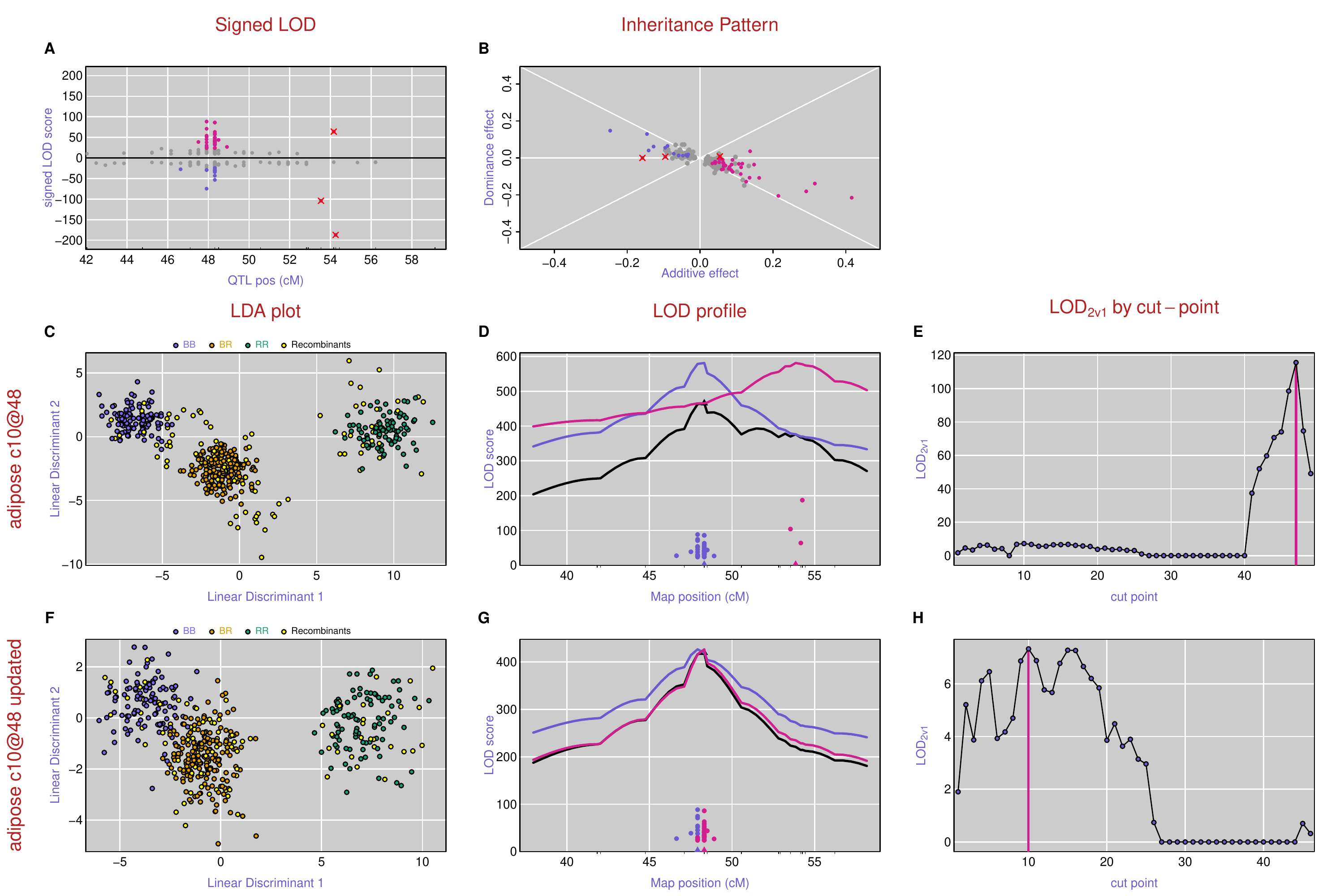}}

\caption{Effect of omitting three transcripts from the chromosome 10 \emph{trans}-eQTL
  hotspot on the analysis results. \textbf{A}: Signed LOD scores, with
  transcripts having LOD $\ge$ 10 highlighted. The three transcripts
  to be omitted are indicated with X's. \textbf{B}: Scatterplot of
  dominanance versus additive effects, with transcripts having LOD
  $\ge$ 10 highlighted. The three transcripts to be omitted are again
  indicated with X's. \textbf{C-E}: LDA and likelihood results, as in
  Figures 3 and 5. \textbf{F-H}: LDA and likelihood results with the
  three transcripts omitted.}

\end{figure}

\clearpage

\begin{figure}[p]
\centerline{\includegraphics[width=\textwidth]{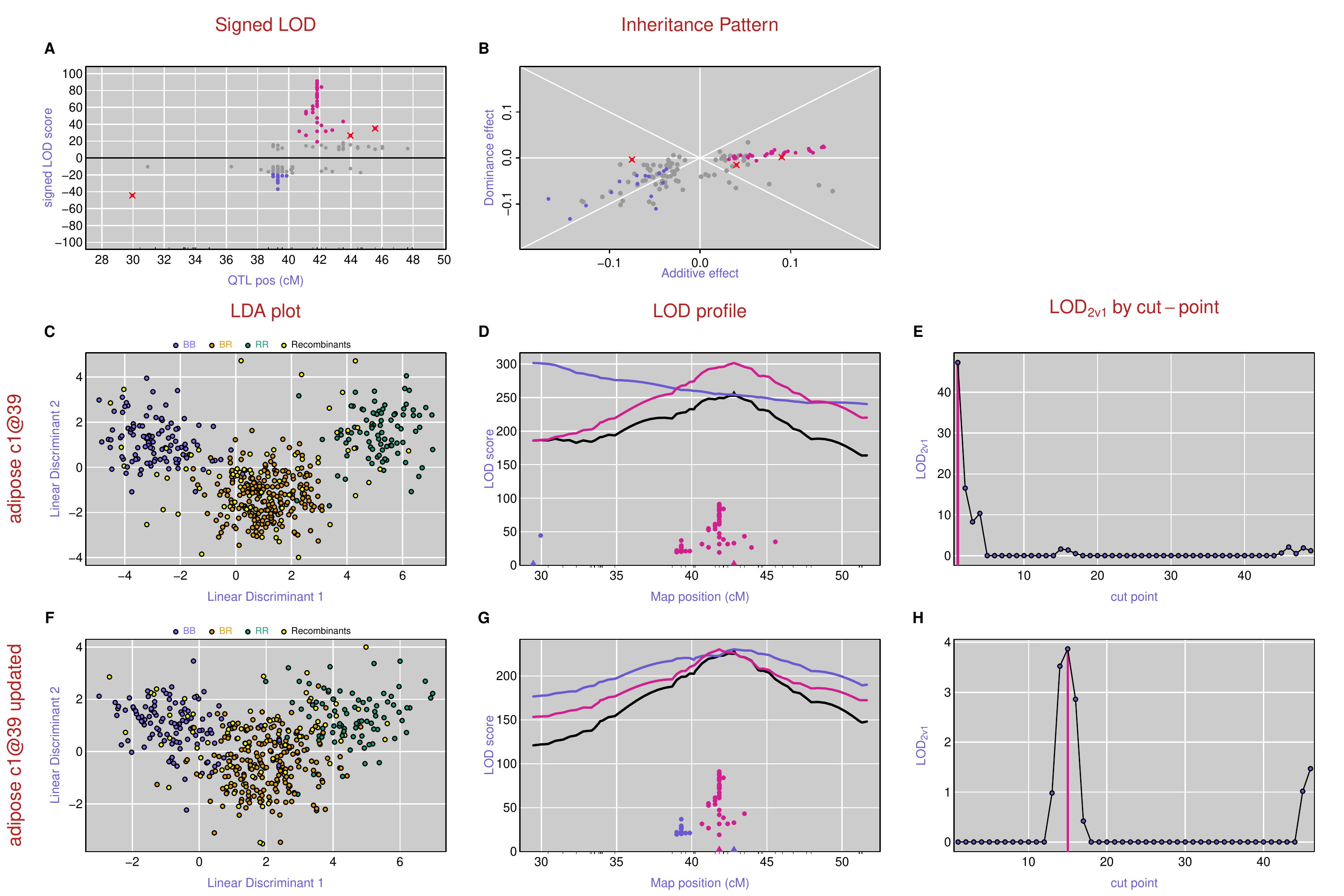}}

\caption{Effect of omitting one transcript from the chromosome 1 \emph{trans}-eQTL
  hotspot on the analysis results. \textbf{A}: Signed LOD scores, with
  transcripts having LOD $\ge$ 10 highlighted. The transcript
  to be omitted is indicated with an X. \textbf{B}: Scatterplot of
  dominanance versus additive effects, with transcripts having LOD
  $\ge$ 10 highlighted. The transcript to be omitted is again
  indicated with an X. \textbf{C-E}: LDA and likelihood results, as in
  Figures 3 and 5. \textbf{F-H}: LDA and likelihood results with
  one transcript omitted.}
\end{figure}

\clearpage

\noindent \textbf{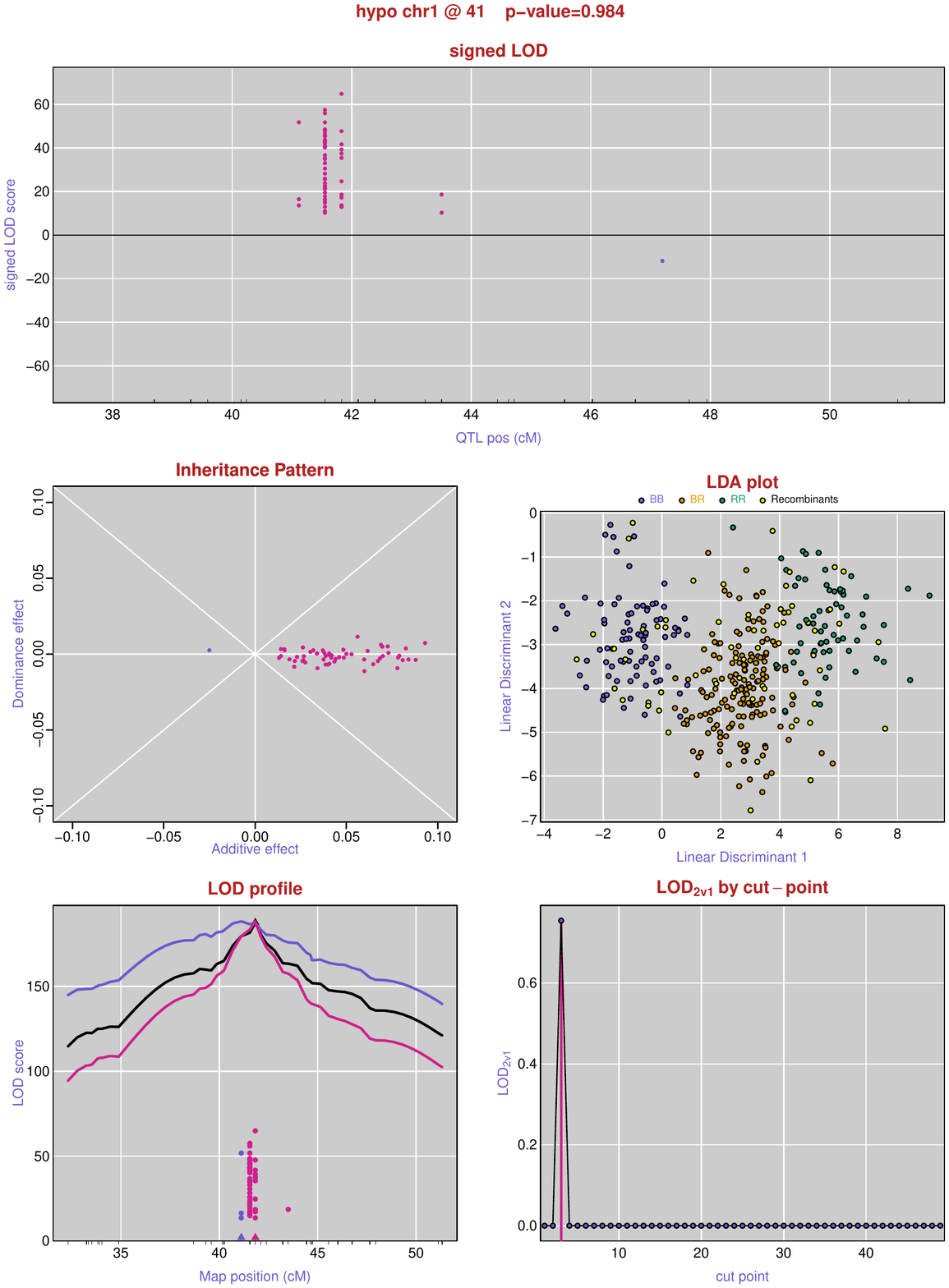} \enspace
A 35-page PDF with the results as in Figures 2, 3, and 4, for all 35
\emph{trans}-eQTL hotspots identified. (Available separately.)

\end{document}